\documentclass{iopart}
\usepackage{epsfig}  
\begin{document}

\font\schwell=schwell at 12pt
\renewcommand{\l}{\schwell L\/}

\title{Short-Lived $\mbox{\boldmath{$\phi$}}$ Mesons}

\author{Laura Holt\footnote[3]{email: hola01@stcloudstate.edu} 
and Kevin Haglin\footnote[4]{email: haglin@stcloudstate.edu}
}

\address{Department of Physics, Astronomy and Engineering Science,
Saint Cloud State University, 720 Fourth Avenue South,
St. Cloud, MN 56301, USA}

\begin{abstract}
We use effective hadronic field theory to study in-medium properties of
the $\phi$ meson.  The dominant decay channels $\phi\to\/K^{+}K^{-}$
and $\phi\to\pi\rho$ are modeled using an SU(3) chiral Lagrangian with
normal and abnormal parities.
The $\phi$ self-energy is approximated to one-, two-, and even three-loop
order in the strong coupling.   Effects of modified 
spectral functions for $\pi$, $K$, $\rho$ and $\phi$ are also included
in the calculation.   This allows us to study in-medium decay of phi mesons 
into (in-medium) kaon-pair daughters and (in-medium) $\pi\rho$ pairs.  The 
results point to the possibility of rather short-lived $\phi$'s, short 
enough to decay inside the fireball in relativistic heavy ion collisions. 
Implications relevant for the NA49 and NA50 experimental results, recently 
dubbed the ``$\phi$ puzzle'', are discussed.
\end{abstract}




\section{Introduction}
Vector mesons are expected to undergo significant changes in heated and 
compressed nuclear matter near the QCD phase boundary as compared to their 
vacuum properties\cite{hirschegg00}.   At the very minimum, one expects 
that spectral distributions will be modified by collisions with light 
mesons\cite{kh95}.  This type of effect has already been seen in the 
CERN experiments looking at the low mass dilepton signals\cite{ceres}.  
The data seem to support a picture where, for example, 
the rho meson is nearly completely melted into the 
background\cite{rw00}.  
The omega
meson scattering rates are roughly the same as the rho's, and so it too is
distorted beyond recognition in an invariant mass plot.  The phi 
meson has up to now been considered to be a bit different.  Its vacuum 
lifetime of 45 fm/$\/c$ puts the decays in all likelihood outside the hot 
reaction zone generated in heavy ion collisions.   Any decays, both hadronic
and electromagnetic, would therefore again be governed by vacuum physics.
And yet, its collision rate has been shown to be 
significant\cite{kh95,vk02}.  Consequently, decay rates could
be affected by the medium.

Meanwhile, there are very puzzling experimental results from CERN
which are beyond description.  First, the NA49 experiment measured the 
momentum distribution for the hadronic channel $K^{+}K^{-}$\cite{na49}.  
The inverse slope parameter suggests an effective temperature around 
305 MeV.  Next, the NA50 experiment reports a dimuon signal whose momentum 
distribution carries an inverse slope parameter of 228 MeV\cite{na50}.  
Given that there are slightly different kinematical ranges covered
by the two experiments, there is some possibility of different effects 
playing roles.  
But we don't expect the effects are different enough to warrant different
physics.  Our goal is therefore to study a hadronic fireball in a common 
framework and to look for a reasonable suggestion for such apparent
temperatures.  If one says that the kaon results are probably indicating 
strong flow, then there is no reason for the dilepton results to be free from 
flow.  Since the lifetime of the phi is long, the flow should be affecting 
both signals consistently.  But this is not what the data are showing, which
is indeed a challenge for theory that one is calling the phi puzzle.  
Model calculations taking medium effects into account have begun
to appear\cite{pf01,pkl02}.

\section{$\boldmath{\mbox{$\phi$}}$ Meson Self Energy}

The essential quantity for studying finite temperature effects on
the vector mesons is the self energy, which is related to the inverses
of the bare and full propagators.  Complete spectral behavior is also 
obtainable from the components of the self energy.  To begin, we must identify 
the relevant hadronic degrees of freedom as well as a model for their 
interactions.
The strangeness content of the $\phi$ dictates that kaons be included
in the model, and also $K^{*}$(892).  Of course, the $\phi$ will interact
with light unflavored mesons $\pi$ and $\rho$, as well.  This argues for
an SU(3) chiral Lagrangian.  We thus start with the nonlinear
sigma model
\begin{eqnarray}
{\cal\/L\/} & = & 
{F_{\pi}^{\,2}\over\/8}\,\partial_{\mu}U\/\partial^{\,\mu}\/U^{\dag},
\end{eqnarray}
where $U = \exp\,(2\,i\phi/\/F_{\pi})$, $\phi$ is
the three-flavor pseudoscalar mesons nonet
and $F_{\pi}$ = 135 MeV is the pion decay constant.
Vector and axial vector mesons are introduced through the
chiral covariant derivative $\partial_{\mu}U\to{D}_{\mu}U
= \partial_{\mu}U - igA_{\mu}^{L}\,U + igU\,A_{\mu}^{R}$, and
the left- and right-handed vector fields $A_{\mu}^{R}$ and
$A_{\mu}^{L}$ are linear combinations
of physical vector and axial vector fields.
Kinetic energy terms for spin-1 fields are added as well as
generalized mass terms for $A_{\mu}^{L}$ and $A_{\mu}^{R}$.  Then, the axial 
vector fields are gauged away leaving to lowest order a set of interactions for
light plus flavored pseudoscalar and vector mesons.  The interactions
are compactly written as\cite{khcg01}
\begin{eqnarray}
{\cal L\/}_{\rm\,int} & = & i\,g\,{\rm\/Tr\,}\left(\rho_{\mu}\/
\left[\partial^{\mu}\phi,\,\phi\right]\right) - 
{g^{\,2}\over\/2\/}{\rm\/Tr\/}\left(
\left[\phi,\,\rho^{\mu}\,\right]^{\,2\/}\right)
+ i\,g\,{\rm\/Tr\/}\left(\partial_{\mu}\rho_{\nu}\left[\rho^{\mu},\,\rho^{\nu}
\,\right]\right)
\nonumber\\
& \ & + {g^{\,2}\over\/4\/}
{\rm\/Tr\/}\left(\left[\rho^{\mu},\,\rho^{\nu}\right]^{\,2\/}\right),
\end{eqnarray}
where $\rho^{\mu}$ is the nonet of vector mesons.
We use the SU(3) symmetry  to dictate the form of the interactions, but
then allow individual coupling strengths to be fixed by appealing
to data.  Since the $\phi$ has a nonnegligible decay branch to
$\pi\rho$, we must also model the abnormal parity interactions.  
We write these as
\begin{eqnarray}
{\cal\/L\/}_{\phi\pi\rho} & = & 
g_{\phi\pi\rho}\,\epsilon_{\mu\nu\alpha\beta}\,\partial^{\mu}\phi^{\nu}
\partial^{\alpha}\vec{\rho}^{\,\beta}\cdot\vec{\pi},
\end{eqnarray}
where $\vec{\rho}^{\,\mu}$ and $\vec{\pi}$ are the complex rho meson
and pion fields.
Empirical constraints are again used to fix the coupling constant.

\begin{figure}[!t]
\begin{center}
\epsfig{file=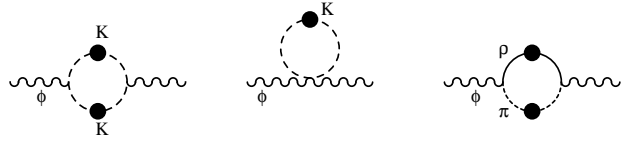,width=4.0cm}
\hspace{0.5cm}
\end{center}
\caption{Self energy graphs considered here.  One-loop correspond
to graphs without blobs, two- and three-loop contributions
include one or two blobs, respectively.  Physically, blobs
are themselves hadronic loop effects.}
\label{feynmand}
\end{figure}

One-loop topologies contribute to the self energy in ways
that correspond physically to thermal adjustments
to the pole mass and to on-shell decays.  We include kaon
bubble and tadpole graphs, and a $\pi$--$\rho$ loop 
(see Fig.~\ref{feynmand}, but for now remove the higher order effects 
which are indicated by blobs).
These one-loop effects have been thoroughly studied in the
literature and turn out to be rather small\cite{khcg94}.  At two-loop
order, the contributions are again thermal adjustments to the
pole mass, off-shell particle decays, and most importantly,
scattering processes such as $\phi+\,K\to\phi+\,K$, 
$\phi+\,K^{*}\to\pi+\,K$, and others.  If one of the internal
kaon lines in the kaon-bubble graph of Fig.~\ref{feynmand} is dressed
with a $K^{*}$--$\pi$ loop, this would then correspond to a 
two-loop contribution
which has dominant influence on the imaginary part of the
self energy.  It corresponds physically to the scattering
process $\phi+\pi\to\,K\,+K^{*}$, as well as other possibilities
for 2$\to$2 body scattering.  Then at 
three-loop order there are many possibilities.  Both internal lines
could be dressed, vertex corrections can be made, etc.  Depending 
on the specific kinematics, the diagrams correspond to  
off-shell decays, off-shell scattering, and even three-body decays.
The cases of particular interest here are the off-shell decays
$\phi\to\,K^{+}K^{-}$ and 
$\phi\to\,\pi\rho$, where {\it\/all} particles are off shell.

\section{Spectral Functions}

The important quantity which is needed for an
assessment of a particle's response to the medium is
the spectral function.  It depends on the real and imaginary
parts of the self energy as follows (for the $\phi$)
\begin{eqnarray}
\rho(M) &=& {1\over\pi}\,{-{\rm\/Im}\,\Pi\over (
M^{2}\,-m_{\phi}^{2}\,-{\rm\/Re}\,\Pi)^{2} + 
({\rm\/Im}\,\Pi)^{2}}.
\label{sfunctions}
\end{eqnarray}
Notice that transverse and longitudinal excitations have not been
distinguished.
Since we are interested in two-, and even three-loop effects,
we will make some simplifying assumptions.  We will
suggest that to a good approximation it is appropriate to
absorb the real part of the self energy into the mass, hence
$\rm\/Re\Pi\approx\,0$.  The imaginary part can be shown to be
directly related to the rate of decays plus collisions, plus
various absorption rates\cite{aw83}.
The rate which dominates in the hot and dense system is the
collision rate.  We will therefore use
\begin{eqnarray}
{\rm\/Im}\Pi & = & -\omega\,\Gamma^{\scriptstyle\rm\/coll}.
\end{eqnarray}

For the general scattering process $\phi + b \to 1 + 2$, the
scattering rate from kinetic theory is
\begin{eqnarray}
d\,\Gamma^{\scriptstyle\rm\,coll} & = &
{g_{a}\,g_{b}\over\,n_{\phi}}
{d^{3}\,p_{\phi}\over(2\pi)^{3}\/2E_{\phi}}\,f_{\phi}
{d^{3}\,p_{b}\over(2\pi)^{3}\/2E_{b}}\,f_{b}
{d^{3}\,p_{1}\over(2\pi)^{3}\/2E_{1}}\,(1+f_{1})
\nonumber\\
& \ & 
\times{d^{3}\,p_{2}\over(2\pi)^{3}\/2E_{2}}\,(1+f_{2})
|\bar{\cal\/M\,}|^{2}\,(2\pi)^{4}\,\delta^{4}(
p_{\phi}+p_{b}-p_{1}-p_{2}).
\end{eqnarray}

In a Boltzmann approximation and an $s$-channel resonance
picture, the collision rate for the
process $a+b\to\/1+2$ can be simplified to
\begin{eqnarray}
\Gamma^{\scriptstyle\rm\/coll}_{a} & = &
{T\,g_{b}\over\/8\pi^{2}m_{a}^{2}{\cal\/K\/}_{2}(m_{a}/T)}
\,\int_{z_{\rm\/min}}^{\,\infty}
\,dz\,\lambda(s,m_{a}^{2},m_{b}^{2})\,{\cal\/K\/}_{1}(z)\,\sigma(s),
\end{eqnarray}
where $g_{b}$ is the degeneracy of species $b$, ${\cal\/K\/}_{i}$ is 
a modified Bessel function of order $i$, $\lambda(x,y,z) = 
x^{2}-2x(y+z)+(y-z)^{2}$, $z = \sqrt{s}/T$ and $z_{\min} = 
[{\rm\/MIN}(m_{a}+m_{b},m_{1}+m_{2})]/T$.

\begin{figure}[!t]
\begin{center}
\epsfig{file=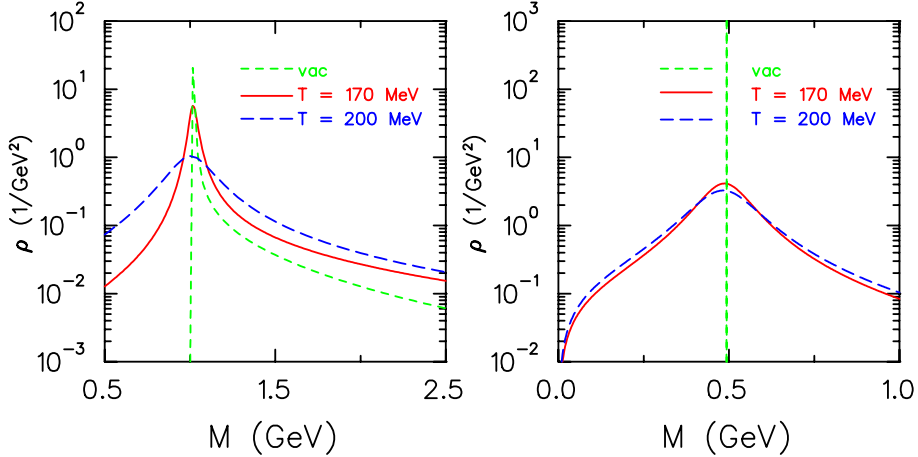,width=6.0cm}
\hspace{0.05cm}
\end{center}
\caption{Spectral function for $\phi$ (left panel) and for the
kaon (right panel).  The dominant effect comes from scattering in the 
medium.}
\label{sfphik}
\end{figure}

Results for the $\phi$ and for kaons are displayed in 
Fig.~\ref{sfphik}.  The striking feature is that near
$T_{c} \approx 170$ MeV, the spectra are quite dramatically
distorted as compared with the vacuum.  Similar pictures can
be generated for the pion and for $\rho$.  Broadening of
$\rho$ has also been well studied in the literature\cite{kh95,kh96}.

Next, we make a comment regarding the physical meaning of the
width of the spectral function.   It is a direct a measure of
the rate for something dynamical to happen.  This could be
a decay, but it turns out to be essentially dominated by elastic and 
inelastic scattering.
When one asks about the production rate of $\mu^{+}\mu^{-}$ or
the rate to decay into $K^{+}\,K^{-}$,
this width is not the relevant one.  However, the in-medium spectral
function can be used to estimate these decay rates\cite{kh04}, as
we discuss below.

\section{Decay Rate} 

On general field theoretic grounds, a resonant hadronic state 
$|R\rangle$ decays into a
two-body final state $|f\rangle = 1 + 2$ with off-shell daughters
at the rate\cite{aw93}
\begin{eqnarray}
{d\/N_{f}\over\/d^{4}x\/d^{4}q\/}
& = & 
{(2J+1)\over\/(2\pi)^{3}}\,
{1\over\/\exp(\beta\/q_{0}) \pm 1}
\,\rho(M)\,2M\/\Gamma^{\rm\/med}_{R\to\/f\/},
\label{invrate}
\end{eqnarray}
where 
\begin{eqnarray}
d\,\Gamma^{\rm\/med}_{R\to\/f\/} & =&
d\,s_{1}\,\rho(s_{1})\,d\,s_{2}\,\rho(s_{2})
\Gamma^{\rm\/vac}_{R\to\/f\/}(M^{2},s_{1},s_{2}).
\end{eqnarray}
The spectral functions for the daughters 1 and 2 come
from Eq.~(\ref{sfunctions}), and 
$\Gamma^{\rm\/vac}_{R\to\/f\/}(M^{2},s_{1},s_{2})$
is the vacuum decay rate with specific invariant masses.
This function is obtainable from knowledge of the interaction Lagrangian.
Integrating Eq.~(\ref{invrate}) over all three momentum
and over all off-shell energies gives the number of decays
per unit time per unit volume.   
From this result, we simply
divide by the number of $\phi$ mesons per unit volume to
arrive at the number of decays per unit time.   This in-medium
decay rate is then inversely related to the lifetime through
\begin{eqnarray}
\tau_{\phi} &=& {1\over\Gamma^{\scriptstyle\rm\/decay}}
= \left\lbrack{1\over\/n_{\phi}}{dN_{f}\over\/d^{4}x}\right\rbrack^{-1}.
\end{eqnarray}
Should the decay lifetime turn out to be on the order of the 
fireball lifetime
or shorter, this would signal that phi mesons indeed decay
inside.  On the other hand, if the medium has little effect
on the lifetime and it remains around 45 fm/$c$, then most likely
the $\phi$'s decay outside.

\begin{figure}[!t]
\begin{center}
\epsfig{file=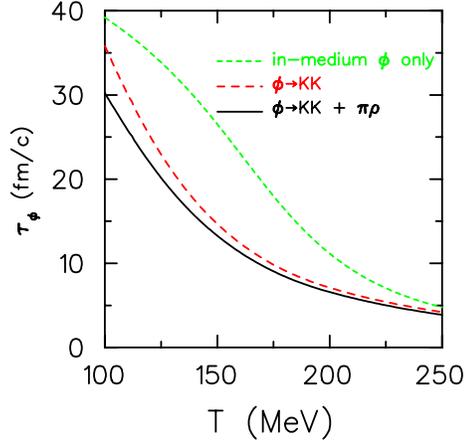,width=6.0cm}
\hspace{0.05cm}
\end{center}
\caption{The in-medium lifetime of the $\phi$ as a function
of temperature.  The solid line takes into account $K^{+}K^{-}$
and $\pi\rho$ off-shell final states, while the short dashed curve
results from off-shell $\phi$ only [and on-shell final states].
Finally, the long-dashed curve accounts only for off-shell
$K^{+}K^{-}$ final state alone.}
\label{tauphi}
\end{figure}

The physical interpretation of the lifetime result, which is
plotted in Fig.~\ref{tauphi}, is the following.  If we including
broadening effects on the $\phi$ spectral function due to collisions 
($\sim$ 40 MeV width near $T_{c}$), but we do not broaden the decay products,
we find the short-dashed curve.  The lifetime in this
case does decrease
at finite temperature to something like 20 fm/$c$ near $T_{c}$.
Again, the broadening here comes physically from two-loop self
energy contributions of scattering the $\phi$ meson with light
mesons.  The medium does however, have an effect also on the
daughters.  We therefore next allow the daughters to scatter,
which means we allow for the daughters to have spectral functions
which are also quite broad due to their respective scattering
rates with pions, rho mesons, etc.  Results are displayed as the
long dashed curve with the off-shell $K^{+}K^{-}$ final state only, and
the solid curve when we also include the off-shell $\pi\rho$ final state.
The specific feature to point out is that the lifetime
of the $\phi$ decreases. It decreases quite rapidly with
rising temperature, dropping to roughly 10 fm/$c$ by 150 MeV,
and 7 fm/$c$ by 200 MeV temperature.

We suggest from the model that the in-medium $\phi$ does
decay inside the fireball!  

\section{Implications for Experiment}

A $\phi$ decaying in the medium into $K^{+}K^{-}$ is unmeasurable
since the kaons will most likely rescatter and be lost.  Here,
lost means that it will not be possible to
reconstruct the parent $\phi$.  However, the dilepton branch 
$\mu^{+}\mu^{-}$, will
not rescatter.  Those muon pairs will escape the system and
be identifiable as having come from the $\phi$.  Those muon pairs
should exhibit the in-medium spectral properties, namely a
collision broadened invariant mass distribution 
of several tens of MeV.  Furthermore, the dilepton
signal, being an increasing function of temperature, ought
to be dominated by the highest temperature, when flow has not
had a great deal of time to build up.

The reconstructed kaon pairs, which are identified as having come
from $\phi$, will be post-freezeout decays emerging from a stage 
when the flow has had sufficient time to build up.  Experimental
indications are that a reasonable value for the flow
velocity is roughly half the speed of light, or greater.
The in-medium spectral behavior of $\phi$, $\rho$, $K$ and
$\pi$ seem to point to this as a possibility.   Therefore,
we next look at the experimental data from CERN in this
picture to extract temperatures and flow values.

To do that we take the Siemens-Rasmussen formula for radial flow\cite{sr79} 
and do a minimum $\chi$-square fit to extract a effective
temperatures and radial flow velocities for the experimental
data.   In this picture, the local rest frame is assumed
to exhibit an equilibrium momentum distribution while 
collectively moving radially outward with speed $v$.  If we Lorentz
transform back to the fireball rest frame we find
\begin{eqnarray}
{d^{2}n\over\/m_{t}\/dm_{t}\/dy\/}
& =& 
{e^{-\gamma\/m_{t}/T}\over(2\pi)^{2}}\left\lbrack
\left(\gamma\/m_{t} +T\right){{\rm\/sinh}(\alpha)\over\alpha}
-T{\rm\/cosh}(\alpha)\right\rbrack,
\end{eqnarray}
where $\gamma = 1/\sqrt{1-\beta^{2}}$ and $\alpha =
\gamma\beta\/|\vec{p\,}|/T$.

\begin{figure}[!t]
\begin{center}
\epsfig{file=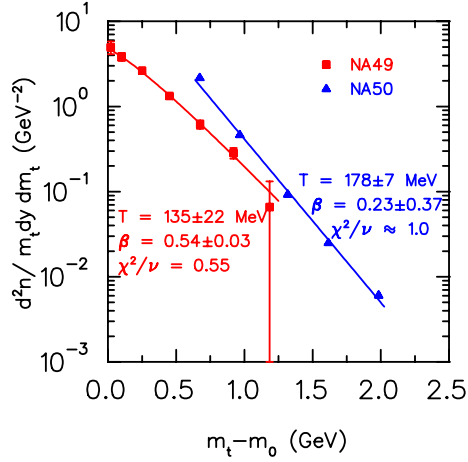,width=6.0cm}
\hspace{0.05cm}
\end{center}
\caption{Inverse slope parameters and flow velocities coming
from a minimum $\chi$-square fit to the NA49 and NA50 data.}
\label{newfit}
\end{figure}

The NA49 and NA50 data are then fit to this functional
form.  Results are shown in Fig.~\ref{newfit}
and seem to indicate first that the kaon spectrum is
consistent with low temperature ($T = 135$ MeV) but
high flow ($\beta = 0.54$).  This is consistent with near
freezeout behavior.  And second, the results indicate the muon pair 
spectrum is
consistent with smaller flow ($\beta
= 0.23$, albeit with very large uncertainty) and a higher 
temperature ($T = 178$ MeV).   This scenario fits with
the model, and fits with the experimental results.
 
\section{Conclusions}

We have applied an effective Lagrangian to describe the dynamics of
$\phi$ mesons in a hot and dense hadronic fireball.  The
formalism attempted to include effects beyond one-loop (on-shell decays),
to include collision broadening effects of the $\phi$ (two-loop
contributions) and even three-loop effects (off-shell decays).  The higher
order effects correspond physically to broadened spectral functions for
not only $\phi$, but also for all the daughter particles
$K^{+}K^{-}$ and $\pi\rho$.  We find that the in-medium
decay rate jumps from its vacuum value of roughly 4 MeV to
nearly 40 MeV at high temperature.  This corresponds to an
in-medium lifetime at high temperature of roughly 5 fm/$c$.  Therefore
we conclude that the $\phi$ meson is likely to decay inside the
hadronic fireball.

This suggestion implies the following scenario for possibly
resolving the NA49 and NA50 ``phi puzzle''.  At high temperatures
($T\approx T_{c}$) the phi lifetime is short, and it will
decay, both into $K^{+}K^{-}$ and $\mu^{+}\mu^{-}$.  The daughter
kaons will be reabsorbed, while the dileptons will reach the detector.
As the system expands and cools, the lifetime for the phi increases
as medium effects are diminished.  The $\phi$ mesons that decay
near the freezeout surface and beyond have returned to vacuum
behavior.  Therefore the two-kaon distributions are expected to
have the free-space width.   The prediction in this
calculation, if there is one, is that the dilepton signal will
show an in-medium spectral function broadened by collisions
while the hadronic signal will show free-space behavior.  Finally,
the apparent branching ratio of the dilepton channel will
increase by something like a factor of 2--5.  This feature is
currently being studied quantitatively\cite{tba}.

\section*{Acknowledgments}
This work has been supported in part by the National Science Foundation
under grant number PHY-0098760.

\end{document}